\documentclass[%
 reprint,
 amsmath,amssymb,
 aps,
]{revtex4-2}

\usepackage{graphicx}
\usepackage{dcolumn}
\usepackage{bm}
\usepackage{diagbox} 
\usepackage{xcolor}
\usepackage{color}
\usepackage[normalem]{ulem}

\begin{document}


\title{Quantifying many-body contributions to depletion forces}

\author{Gabriel P\'erez-\'Angel}
\email{gperez@cinvestav.mx}
\affiliation{Departamento de F\'isica Aplicada, Cinvestav Unidad M\'erida, AP 73 Cordemex, M\'erida, 97310, Yucat\'an, Mexico}

\author{Marco A. Ram\'irez-Gu\'izar}
\email{ma.ramirezguizar@ugto.mx}
\affiliation{Divisi\'on de Ciencias e Ingenier\'ias, \\
Universidad de Guanajuato, Le\'on, 37150, Mexico}

\author{N\'estor M. De Los Santos-L\'opez}
\email{nestor92sl@gmail.com}
\affiliation{Divisi\'on de Ciencias e Ingenier\'ias, \\
Universidad de Guanajuato, Le\'on, 37150, Mexico}

\author{Jos\'e M. M\'endez-Alcaraz}
\email{josem.mendez@cinvestav.mx}
\affiliation{Departamento de F\'isica, Cinvestav, Av.\ IPN 2508, Col.\ San Pedro Zacatenco, Gustavo A.\ Madero, 07360, Ciudad de M\'exico, Mexico}

\author{Ram\'on Casta\~neda-Priego}
\email{ramoncp@fisica.ugto.mx}
\affiliation{Departamento de Ingenier\'ia F\'isica, Divisi\'on de Ciencias e Ingenier\'ias, \\
Universidad de Guanajuato, Le\'on, 37150, Mexico}

\date{\today}

\begin{abstract}

Effective interactions inherently encompass many-body effects that appear unified. Analyzing these in reverse, that is, separating them into contributions from pairs, triples, or larger groups, is typically intricate and seldom pursued. However, this could offer essential insights into the structural architecture of complex systems, such as soft materials. This contribution tackles this issue employing a new simulation-based approach to accurately determine effective interactions [J. Chem. Phys. {\bf 157}, 074903 (2022)]. This scheme is applied within the semi-grand canonical ensemble to calculate depletion forces in asymmetric binary mixtures of hard spheres. The approach is sufficiently sensitive to assess the effects of higher-order terms in the depletion potentials between large particles. Previous research has primarily focused on the role of small particles, and this work expands on these findings. However, understanding the contributions of large particles on the effective forces, particularly beyond the dilute limit, remains challenging and is not yet fully grasped, leading us to primarily focus on exploring this topic. Within the range of particle concentrations examined here, while maintaining a constant chemical potential for smaller particles, we have observed that the concentration of larger particles has no impact on the entropic potential between large colloids, as long as the size ratio remains below $q=0.15$. This confirms a long-established prediction founded on purely geometric considerations, which we have confirmed by means of direct observation. In contrast, for mixtures with less size asymmetry, such effects become considerably influential. Specifically, we have conducted an in-depth examination of scenarios with size asymmetries of $q=0.45$ and $0.60$. Our results for the depletion forces were directly compared with those obtained from the integral equation theory, which has also enabled us to improve and refine the approaches involved.

\end{abstract}


\maketitle


\section{\label{introduction} Introduction}

The importance of many-body interactions in atomic physics was recognized
in the pioneering work of Axilrod and Teller~\cite{axilrod_JCP_1943}, and in the early simulations in silicon done by Stillinger and
Weber~\cite{stillinger_PRB_1985}. For atomic and molecular systems, it has been
noticed that in many cases, the use of these many-body interactions is fundamental for the convergence of simulation results to experimental values \cite{bukowsky_JCP_2001}, reaching even situations in which, for some geometries, the two-body interactions become null and the dynamics is governed completely by three-body and higher-order forces~\cite{yan_PRA_2021}. The use of these many-body terms is common enough that, in a simply parameterized form (3-body angles, 4-body dihedral angles), they are already included in many computational packages \cite{gromacs_manual, lammps_manual}. In basically all these cases, it is assumed that the forces are calculable in some way (perturbation theory in Ref. ~\cite{axilrod_JCP_1943}, for example), and these known interactions are just added to the fundamental forces in a molecular simulation.

In colloidal-like dispersions and, more generally, soft matter systems, 
there is usually no \emph{a priory\/} way to include these
many-body contributions to an effective potential. As will be discussed shortly, ingenious efforts have been made to disentangle many-body interactions, particularly three-body potentials, in these mesoscopic aggregates. For example, in some cases, direct measurements of three-body potentials have
been carried out for equilateral and isosceles triangles,
using density functional theory (DFT)~\cite{goulding_PRE_2001,frischknecht_JCP_2011}, and
in direct Monte Carlo simulations~\cite{zhu_JP:CM_2003},
or by combining molecular dynamics simulations with DFT and comparing the results with the DLVO-like description in charged colloidal suspensions ~\cite{lowen_JP:CM_1998}.
Another possibility is to generate second and third virial coefficients and extract from those the three-body interactions,
either in simulations~\cite{ashton_JCP_2014}, or with
purely theoretical methods~\cite{santos_JCP_2015}.
Finally, a different approach is that of calculating the total contribution
of many-body forces, independently of geometry, to some observable measured in equilibrium. This has been done
using, for example, the Poisson-Boltzmann equation to determine effective potentials with the inclusion of many-body contributions in charge-stabilized colloidal dispersions and then comparing them with
the effective pair potential obtained from a DLVO-like picture~\cite{dobnikar_NJP_2006}. 
	
As noted above, it remains a theoretical and technical challenge to explicitly quantify the effects of many-body contributions on the effective potential between macromolecules at a finite concentration. To make further progress in this direction, and without loss of generality, our contribution focuses on a particular case of many-body effects that influence effective potentials in soft matter, commonly referred to as depletion interactions, which can be accurately determined even beyond the diluted limit by applying the so-called contraction of the bare forces (CBF) scheme; see, e.g., Refs. \cite{delossantos-lopez_JCP_2022,delossantos-lopez-PA-2024}.

The depletion forces were introduced by Asakura and Oosawa (AO) \cite{asakura_JCP_1954,asakura_JPS_1958} 
(and independently in \cite{vrij_PAC_1976})
to explain some phenomena of the aggregation of large biomolecules
(in the colloidal size regime), immersed in a solvent populated by much smaller polymeric coils (or proteins) \cite{oosawa_JCP_2021}. The physics behind the depletion interaction comes from the fact that in these types of mixtures, the colloidal particles, usually assumed to be spherical,
present an \emph{excluded volume\/}
to the polymer coils; that is, there is a spherical shell surrounding the colloid where the centroid of the polymeric coil (also taken as spherical) cannot enter.
In this model, when two colloidal particles are close enough, their excluded
volumes overlap and, as a consequence, the total excluded volume
decreases and the volume accessible to the polymers increases.
This results in a decrease in the free energy of the mixture that is large enough to generate an attractive force between the colloids, on the order of the thermal energy. It is now common to name colloids the \emph{depleted\/}
particles and polymers the \emph{depletants} \cite{oosawa_JCP_2021}. 
It should also be noted that the depletion interaction, when the particles involved
are taken as hard spheres, is purely entropic. 

The original AO model assumed full interpenetrability of the polymers, so they behaved among themselves as an ideal gas. In that case, and assuming the infinitely diluted limit for the colloids, the interaction potential can be calculated analytically \cite{gast_JCIS_1983,lekkerkerker_2011}. For situations where the colloidal dispersion has a finite density, Dijkstra and coworkers \cite{dijkstra_PRL_2002,dijkstra_PRE_2006} have found a way to construct the effective Hamiltonian, but always assuming an ideal gas of polymer coils.

It should be clear that the ideas behind the AO model generalize to situations involving large and small nonoverlapping colloids with finite densities for both species. For such more realistic cases, a long sequence of works has found approximations to the complete depletion potential for binary mixtures of hard spheres. Several of them involve density functional theory (DFT) \cite{Roth2000,dijkstra_PRE_2006}, integral equation theory \cite{alcaraz-klein,ramon-ro-al-2,erik,santos_JCP_2015, Heinen18}, and, of course, recent computer-based schemes that allow accurate calculation of depletion forces \cite{delossantos-lopez_JCP_2021,delossantos-lopez_JCP_2022,delossantos-lopez-PA-2024}.

Using the CBF approach and integral equation theory for effective interactions, in this contribution, we demonstrate that in cases of significant differences in the sizes of the mixture components, where the diameters of the smaller components are $q\le 0.15$ of the larger ones, the many-body effects are negligible. This allows for simplifying assumptions that additive pair interactions adequately describe the total depletion interaction, as commonly proposed in theoretical models. Notably, in these scenarios, the depletion forces affecting the larger mixture component remain constant, regardless of its density, provided that the chemical potential of the smaller component remains unchanged. In contrast, mixtures with size ratios of $q=0.45$ and $q=0.60$ exhibit a clear influence of the density of the larger component on the effective forces, despite maintaining a fixed chemical potential for the smaller component. Our analysis reveals that in these less asymmetric mixtures, particularly within attractive regions, depletion forces can be accurately modeled by low-degree polynomials. The coefficients of these polynomials can then be used to predict depletion forces in the limit of infinite dilution or to interpolate them at other density points not included in our computational simulations. Interestingly, during the determination of the depletion forces, our CBF scheme is able to explicitly distinguish between the two-, three-, and higher-order many-body contributions in a manner that allows the resulting depletion potential to quantify the relevance of each contribution. This is certainly the most significant result of this work.

\section{\label{DEP}When three-, four-, and many-body interactions become important in determining effective potentials}

In Statistical Physics, most interaction models start as summations of interactions by pairs in simple systems with very large numbers of degrees of freedom. When considering effective interactions, however, one is trying to eliminate some of these degrees of freedom, assumed to be no longer needed for the
description of the effective model. This elimination (or ``tracing out'') of the now Irrelevant Degrees of Freedom (IDFs) proceeds through the taking of some
averaged approach, where the IDFs are allowed to reach equilibrium in configuration space, according to the original Hamiltonian \cite{likos_PR_2001}, with the proviso that the remnant degrees of freedom --- Effective Degrees of Freedom (EDFs), from now on --- are kept fixed. In other words, we allow the IDFs to equilibrate
in the fixed field of a given configuration of the EDFs, and from here,
one gets a value for the potential for EDFs in that given configuration.

This procedure is thoroughly described in a formal manner in \cite{likos_PR_2001} for the Canonical Ensemble $NVT$ and in \cite{dijkstra_PRL_1998,dijkstra_PRE_1999,dijkstra_JP:CM_1999}
for the Semi-Grand Ensemble $\mu VT$, with fixed chemical potential, $\mu$, for the depletants, and fixed $N$ for the depleted particles. In both
situations, one immediately finds that the tracing out of the IDFs introduces three-, four-, and higher order terms in the potential for the EDFs. In practice,
it is known that for any of these descriptions, the resulting total effective potential energy
for a system of $N$ relevant particles should be written as,
\begin{eqnarray}
	U({\bf r}_1, {\bf r}_2, {\bf r}_3, ... {\bf r}_N) &=&
	\sum_i U^{(1)} ({\bf r}_i) +
	\frac{1}{2!} \sum_{j \ne i} U^{(2)} ({\bf r}_i, {\bf r}_j)\nonumber \\ &+& 
	  \frac{1}{3!} \sum_{k \ne j \ne i} U^{(3)} ({\bf r}_i, {\bf r}_j,  {\bf r}_k) \nonumber \\
	&+& ...
	+ U^{(N)} ({\bf r}_1, {\bf r}_2, {\bf r}_3, ..., {\bf r}_N),
	\label{full-potential}
\end{eqnarray}
where the coefficient $1/n!$ before the summation of equation $U^{(n)}$ corrects the overcounting generated by the permutations of the
arguments. Notice that, for convenience, the last term is given with
no summations and just the preset ordering of its variables.
It is not immediately obvious how to disentangle contributions to the potential from
different levels, that is, how to determine the distinction between, for instance,  $U^{(3)} ({\bf r}_1, {\bf r}_2,  {\bf r}_3)$ 
and 
$U^{(2)} ({\bf r}_1, {\bf r}_2) + U^{(2)} ({\bf r}_2, {\bf r}_3) + U^{(2)} ({\bf r}_3, {\bf r}_1)$.
We will just assume that some well-defined set of rules allows for this separation. This is needed so that, for example, one can distinguish between
$U({\bf r}_1, {\bf r}_2, {\bf r}_3, ... {\bf r}_N)$ and 
$U^{(N)} ({\bf r}_1, {\bf r}_2, {\bf r}_3, ..., {\bf r}_N)$ in Eq.~(\ref{full-potential}).

It is well known that for translational invariant spaces, the potential given in Eq.~(\ref{full-potential}) can be written in terms of relative positions with respect to some chosen particle, say the $i$-th particle, which thus is shifted to the
origin of coordinates. Eq. (\ref{full-potential}) can be written as
\begin{eqnarray}
	U({\bf r}_1, {\bf r}_2, {\bf r}_3, ... {\bf r}_N) &=&
	\frac{1}{2!} \sum_{j \ne i} U^{(2)} ({\bf r}_{ij}) \nonumber \\
	&+& \frac{1}{3!} \sum_{k \ne j \ne i} U^{(3)} ({\bf r}_{ij}, {\bf r}_{ik}) \nonumber \\
	&+& ... 
	 + U^{(N)} ({\bf r}_{i1}, {\bf r}_{i2}, {\bf r}_{i3}, ..., {\bf r}_{iN}),
	\label{full-potential2}
\end{eqnarray}
with the usual definition ${\bf r}_{ij} = {\bf r}_j - {\bf r}_i$.
In this case, the potential $U^{(n)}$ just needs $n-1$ arguments because there is no external potential $U^{(1)}$. Again, the term $U^{(N)}$ is left without summations, keeping the original ordering of its arguments, and with the understanding that the null displacement ${\bf r}_{ii}$ is left out.

From now on, we consider the case of a binary mixture of hard or semi-hard spheres. Then, the resulting effective forces between the large particles (\emph{depleted\/}) are the well-known depletion forces, where the small particles (\emph{depletants\/}) are the source of the depletion forces and will be traced out from the description.
In this particular context -- indeed, within any system having short-range potentials that permit only minor overlaps, --- one should notice from the beginning that Eqs.~(\ref{full-potential}) and (\ref{full-potential2}), even if completely general, include too many terms. Thus, unless very long-range interactions and/or correlations appear in the depletants, there is no reason for the depleted particles to have effective interactions with more than one, and possibly two, layers
of their nearest neighbors. In that sense, and for large enough $N$, 
the last term in Eqs.~(\ref{full-potential}) and (\ref{full-potential2}), 
and many of its preceding ones, are zero. 

A very clear and noteworthy example
of this truncation has been known for some time,
where the depletion forces that appear in a mixture of hard spheres 
are free \cite{dijkstra_JP:CM_1999} 
or almost free \cite{dijkstra_PRL_1998,dijkstra_PRE_1999} from many-body effects, that is, become 
\emph{exact or almost exact interactions-by-pairs\/}
for the mixtures when,
\begin{equation}
	q \equiv \frac{\sigma_s}{\sigma_l} < q_3 = \frac{2}{\sqrt{3}} -1 \approx 0.1547,
\end{equation}
where $\sigma_l$ and $\sigma_s$ are the diameters of the large and small particles, respectively, and $q$ is the size ratio. 

This independence becomes quite evident when the system is described in the Semi-Grand ensemble, where
one finds that the depletion force ${\bf f}_d$ becomes a function of
the chemical potential of depletants, $\mu_s$, and the packing fraction of depleted
particles, $\phi_l$, with the particularity that this dependence on $\phi_l$ vanishes
for $q < q_3$, that is, the depletion force becomes a function of $\mu_s$ only.
The partial packing fractions of the large and small species, $\phi_l$ and $\phi_s$, respectively, are given by
\begin{equation}
\phi_s = \frac{N_s \pi \sigma_s^3}{6V}, \quad
\phi_l = \frac{N_l \pi \sigma_l^3}{6V},
	\label{phi_def}
\end{equation}
with $N_s$, $N_l$ being the number of small and large particles, respectively, and $V$ the volume of the entire system.

This particular result has a simple geometrical origin: If we put together three
large hard spheres in a configuration where they all touch, a small hard sphere will be able to pass through the hole or cavity in the middle as long as $\sigma_s < q_3 \sigma_l$. Or, in other words, below this size ratio, a small particle can
never simultaneously touch three or more large particles. However, it is important to remain cautious and consider two potential scenarios: (a) When all the particles are modeled as hard spheres, the two-body effective potentials are anticipated to nearly encompass the total effective potential energy. This is because the primary contribution to the three-body effect arises from configurations where one small particle interacts with three large particles, which, as demonstrated, do not occur. Nevertheless, there are more intricate arrangements involving three large particles and two or more small particles that might contribute to three-body and higher many-body effects. In this context, it has been stated that effective interactions are \emph{almost exactly\/} described by a two-body interaction. (b) In the Asakura-Oosawa scenario, where the small particles do not interact (indicating that they interact as though they are an ideal gas among each other), such intricate configurations do not exist, as they rely on interactions between small particles. Thus, in this size ratio limit, the effective potentials are \emph{exactly\/} two-body potentials.

This geometric argument can be further extended by considering the situation in which a small particle fits perfectly inside the tetrahedron formed by four large spheres that are in contact with each other. In this case, a small sphere with $\sigma_s < q_4 \sigma_l$, where $q_4 = \sqrt{3/2} -1 \approx 0.2247$, will never touch simultaneously four large spheres, and therefore the effective forces will have two- and three-body components, but no higher-order terms. 

The previous geometrical arguments extend approximately to semi-hard spheres. As long as the potential grows sufficiently fast upon approaching the particles and the temperature is not too large, compared to the potential energy scale, their behavior will reproduce well enough that of real hard spheres \cite{Baez2018}, and the eventual touching of three larger spheres with a smaller one will be extremely unlikely, even if not strictly forbidden.  
The results section will include some discussion of this approximation.

In this work, we address several questions related to the issue of many-body effects on depletion forces in binary mixtures. To achieve this, we use the Semi-Grand ensemble, where $q < q_3$ should give us forces independent of $\phi_l$. 
First, we ask the following question. Is it possible to actually detect the contributions of many-body interactions to the actual depletion forces in these types of mixtures? Second, can the difference between $q < q_3$ and $q > q_3$ be clearly seen in the forces? Third, if so, in what way do many-body effects depend on the packing fractions $\phi_s$ and $\phi_l$? And finally, given that the depletion forces are a function of the interparticle distance, how far in particle separation do the many-body effects go? Every single one of these questions is comprehensively covered here.

\section{Capturing many-body contributions at the level of effective two-body forces}

Applying the method outlined in the Appendix \ref{DF_CBF}, our CBF strategy enables us to identify the effective two-body forces within a binary mixture. Subsequently, it is necessary to evaluate the many-body framework described by Eqs.~(\ref{full-potential}) and (\ref{full-potential2}) in the context of forces. The force acting on the $i$-th large particle can be derived from the potentials specified in Eq.~(\ref{full-potential2}) as follows:
\begin{eqnarray}
	{\bf F}_i(\{{\bf r}_n\})  &=&
	-\nabla_i U(\{{\bf r}_n\})  \nonumber \\
	&=&\sum_{j \ne i} {\bf f}^{(2)}_{ij} ({\bf r}_{ij}) +
	\sum_{k, j \ne i} 
	{\bf f}^{(3)}_{ij} ({\bf r}_{ij},{\bf r}_{ik}) + ...\nonumber \\
	&=&\sum_{j \ne i} f^{(2)}_{ij} ({\bf r}_{ij}) \hat{\bf r}_{ij} +
	\sum_{k \ne j \ne i} 
	f^{(3)}_{ij} ({\bf r}_{ij},{\bf r}_{ik}) \hat{\bf r}_{ij} \nonumber \\ &+& ...,
\end{eqnarray}
where $\{{\bf r}_n\}$ denotes the set containing all ${\bf r}_n$ values. We have considered homogeneity and relocated particle $i$ to the origin.
 
The essential setup at the conclusion reveals that the force acting on each particle remains a cumulative effect of forces directed from all other particles within the colloidal dispersion. Yet, at present, there exist third-order and higher contributions to this particle-particle force. To fully clarify this, consider the force exerted by particle~$j$ on particle~$i$:
\begin{eqnarray}
    {\bf f}_{ij}(\{{\bf r}_n\}) &=& f^{(2)}_{ij}({\bf r}_{ij}) {\bf \hat r}_{ij} +
	\sum_{k \ne (i,j)} f^{(3)}_{ij} ({\bf r}_{ij} | {\bf r}_{ik}) {\bf \hat r}_{ij} \nonumber\\ &+& 
	  \frac{1}{2} \sum_{l \ne k \ne (i,j)} f^{(4)}_{ij} ({\bf r}_{ij} | {\bf r}_{ik}, {\bf r}_{il}) {\bf \hat r}_{ij}+ ...
\end{eqnarray}
In the prior equation, a vertical bar is utilized to distinctly mark the primary argument, ${\bf r}_{ij}$, from the arguments associated with other particles, hereafter referred to as \emph{bystanders\/}. Similar to Eqs.~(\ref{full-potential}) and (\ref{full-potential2}), a $1/(n-2)!$ coefficient is included before $f^{(n)}$ to adjust for overcounting.

Let us now examine the scenario where the effective force between any particle pair, ${\bf f}^{\text{eff}}_{ij}$, can be determined independently of the positions of other particles. Typically, the primary component of this force is represented by ${\bf f}^{(2)}_{12}({\bf r}_{12})$, derived from $U^{(2)}({\bf r}_{12})$. Nonetheless, three-body, four-body, and higher-order interactions also play a role in shaping this effective two-body force. This influence is perceived as a collective effect, requiring the averaging of bystander particle positions across the ensemble considered; practically, we derive these influences by averaging the pertinent forces over the anticipated densities of the third, fourth, or higher-order bystanders surrounding the selected pair $ij$. Consequently, we can express these influences in a form like,
\begin{eqnarray}
	{\bf f}^{\text{eff}}_{ij}({\bf r}_{ij}) &=&
	{\bf f}^{(2)}_{ij} ({\bf r}_{ij}) + 
	  \int d^3{\bf r} \, {\bf f}^{(3)}_{ij} ({\bf r}_{ij} | {\bf r}) \rho^{(3)} ({\bf r}_{ij} | {\bf r}) \nonumber \\ 
      &+& \frac{1}{2} \int d^3{\bf r} \, d^3{\bf r}^\prime \, 
	{\bf f}^{(4)}_{ij} ({\bf r}_{ij} | {\bf r}, {\bf r}^\prime) \rho^{(4)} ({\bf r}_{ij} | {\bf r}, {\bf r}^\prime) \nonumber \\
	&+& ...,
	\label{force-ij}
\end{eqnarray}
where $\rho^{(n)} ({\bf r}_{ij} | {\bf r}, {\bf r}^\prime, ...)$ represents the predicted particle densities located at ${\bf r}$, ${\bf r}^\prime$, and so forth, given that particles $i$ and $j$ are situated at positions ${\bf r}_i$ and ${\bf r}_j$, respectively. It should be noted that Eq. (\ref{force-ij}) defines the effective two-body force influenced by many-body contributions which originate from the same two-particle potential. In essence, the resulting effective two-body force inherently includes many-body influences at the bare interaction level.
Now, these particular densities can be easily written in terms of the multiparticle correlation
functions $g^{(n)}$ \cite{dhont1996introduction, gardiner2009stochastic}
\begin{equation}
	\rho^{(n)} ({\bf r}_{ij} | {\bf r}, {\bf r}^\prime, ...) = 
	g^{(n)} ({\bf r}_{ij} | {\bf r}, {\bf r}^\prime, ...) \rho_0^{n - 2}.
\end{equation}
In a more explicit way, $\rho^{(3)} ({\bf r}_{ij} | {\bf r}) = g^{(3)} ({\bf r}_i, {\bf r}_j, {\bf r}) \rho_0$,
$\rho^{(4)} ({\bf r}_{ij} | {\bf r}, {\bf r}^\prime) = g^{(4)} ({\bf r}_i, {\bf r}_j, {\bf r}, {\bf r}^\prime) \rho_0^2$, etc. Here, $\rho_0$ represents the number density of large particles, given by $\rho_0 = N_l/V$.
Then, Eq.~(\ref{force-ij}) becomes
\begin{eqnarray}
    {\bf f}^{\text{eff}}_{ij}({\bf r}_{ij}) &=& f_{ij}^{\text{eff}}(r_{ij}) {\bf \hat r}_{ij} \approx 
	{\bf f}^{(2)}_{ij} ({\bf r}_{ij}) \nonumber \\ &+&
	\left( 
	F^{(3)}(r_{ij}) \rho_0 + \frac{1}{2} F^{(4)}(r_{ij}) \rho_0^2 + ... \right) {\bf \hat r}_{ij}.
	\label{force-polynomial}
\end{eqnarray}
Therefore, the effective force between two given large particles becomes approximately a power series expansion in terms of the large component number density, with coefficients $F^{(n)}$ that depend on the configuration of the $n$ particles. 

The approximation lies not only in the truncation of the expansion, but also in the fact that the coefficients $F^{(n)}$ still depend on $\rho_0$ through their $g^{(n)}$ arguments. However, one may assume that these $g^{(n)}$ correlation functions can be, in turn, expressed as a power series in the density of particles; for instance, for hard spheres of diameter $\sigma$, it should be possible to write
\begin{equation}
g^{(2)}(r, \rho_0)  = g^{(2)}_0(r) + \rho_0 g^{(2)}_1(r) + \rho_0^2 g^{(2)}_2(r) + ...,
\end{equation}
where $g^{(2)}_0(r) = \Theta(r - \sigma)$, with $\Theta(x)$ the Heaviside function, and $g^{(2)}_1(r), g^{(2)}_2(r)$, etc., are higher-order contributions that incorporate the oscillations found in $g^{(2)}(r)$ for dense fluids. Accepting this expansion, one then sees that the previous power series for the effective forces becomes an approximation only because of the truncation in the number of terms \cite{Kolafa2006,Nijboer1952,Santos2016}.

From Eq. (\ref{force-polynomial}), it becomes clear that the effective two-body force always has many-body contributions ($F^{(3)}, F^{(4)}$, etc.) that, in principle, can be disentangled or extracted from the two-body contribution. Although this is not a trivial or straightforward task, we provide a route from both the CBF and integral equations theory approaches, outlined in Appendix A, to distinguish between the two-, three-, and higher-order many-body contributions in such a way that the resulting effective potential can clearly quantify the relevance of each contribution.

\section{\label{SD} Determination of the chemical potential of the depletants and simulation details}

\subsection{Fixing the chemical potential of the depletants in a Molecular Dynamics simulation}

In contrast to Monte Carlo (MC) methods, there is no fully adopted method for Molecular Dynamics (MD) simulations at fixed chemical potential (Grand Canonical Ensemble) \cite{allen_OUP_2017}. A simple approach is to graft some MC particle insertion-deletion steps into an MD run. Recently, more elaborate algorithms have been developed, such as the possibility of considering fractional particles that ``grow'' into real particles or ``shrink'' until they leave the system \cite{boinepalli_JCP_2003,eslami_JCC_2007}. Another very interesting algorithm considers particles that approach the assembly via an extra spatial dimension,  or leave it in the same way \cite{belloni_JCP_2019}. In this work, basically for convenience, we resort to the old tried method of simply measuring the chemical potential $(\mu)$ of some not too dense particle assemblies, evolving according to some MD algorithm, via the Widom insertion trick \cite{widom_JCP_1963,dijkstra_PRE_1999}. This approach requires several simulations using different parameters, but ultimately yields the relationship between $\mu$ and any special parameter we may want to vary. It is subject, however, to the low efficiency that affects Widom's trick for dense assemblies.

We therefore calculate the curves $\mu_s(\phi_s, \phi_l)$ (in the simulations, we will always use $T = 1 \to \beta = 1$), and invert the relationship to get the different values of $\phi_s$ associated to a given $\mu_s$, for several values of $\phi_l$. In concrete, we calculate $\mu_s$ for several values of $\phi_s$ at $\phi_l = 0$; these initial packing fractions are \emph{reservoir packing fractions\/} for small particles, denoted $\phi_s^{\text{res}}$, and the goal is to find the values of $\phi_s$ that give the same $\mu_s$, but now in the presence of moderately dense populations $\phi_l$ of large particles, at the same temperature $T = 1$. It should be noticed that since at the end we are only interested in the relationship $\phi_s(\phi_s^{\text{res}}, \phi_l)$, we can ignore several contributions to the ideal part of the chemical potential, in particular those related to Planck's constant \cite{frenkel_EPJP_2013}.

\subsection{Molecular Dynamics simulation details}

We simulate binary mixtures of $N_l$ large and $N_s$ small particles, interacting through the modified Weeks-Chandler-Andersen (mWCA) potential
{\footnotesize
\begin{eqnarray}
	u_{ij} (r_{ij})  = \left\{
	\begin{array}{cc}
	\epsilon \left[ 
	\left( \frac{\sigma_{ij}}{r_{ij}}\right)^{12} - 
	2 \left( \frac{\sigma_{ij}}{r_{ij}}\right)^6 + 1\right] 
		&\quad \mbox{if} \quad r_{ij} < \sigma_{ij} 
        \\
		0 & \quad \mbox{otherwise},
	\end{array}
	\right.
	\label{mWCA}
\end{eqnarray}
}
with $\sigma_{ij} = (\sigma_i + \sigma_j)/2$. We perform Molecular Dynamics (MD) simulations in a cubic volume box $V$ with periodic boundary conditions in the canonical ensemble $(NVT)$ at a temperature $k_{B} T/\epsilon = 1.0$. The length, mass and energy scales are set by fixing $\sigma_l = 1$, $m_l = 1$, and $\epsilon = 1$. The mass of a large particle is $m_l$, while the mass of the smaller particles is in the same proportion to it as their volumes. The volume fractions $\phi_s$ and $\phi_l$ have been defined in Eq.~(\ref{phi_def}). We have considered three values for $q$: $q = 0.15$, where we expect to see little ---if any--- effects of the many-body interactions, and $q = 0.45$ and $0.60$, where those interactions should have some effect in the measured forces. The values used for the packing fractions were for the depleted species: $\phi_l = 0$, which provides the values of $\mu_s$ for the reservoir, denoted $\mu_s^{\text{res}}$, and $\phi_l = 0.1$, $0.2$, $0.3$, and $0.4$. For the depletant species in the particle reservoir, we used $\phi_s^{\text{res}} = 0.025, 0.050, 0.075$, and $0.100$.

We start by placing the particles in random positions within the volume, ensuring no overlaps, and the configurations evolve using the Velocity-Verlet algorithm~\cite{allen_OUP_2017}. We use two Bussi-Donadio-Parrinello thermostats \cite{bussi_JCP_2007} to fix the temperature, one for each species, and run a transient long enough to reach equilibrium. Since the total densities used here are not too large (the largest total packing fraction is 0.5), these transients are short, and there are no issues of dynamical arrest or aging. After the transient, we start to calculate the forces.

In all cases, the measurement of $\mu_s$ has been carried out using Widom's method, and the number of particles used has always been large enough to minimize expected errors ${\cal{O}}(1/N_s)$ ($N_s > 16,000$). From test runs with smaller numbers of particles, we estimate that our values of $\mu_s$ are correct to four significant digits (or more). From the plots of $\mu_s$ \emph{vs\/}. $\phi_s$ for each $\phi_l$, we obtain the values $\phi_s(\phi_l)$ that generate, with numerical precision, the same $\mu_s$ as that of the chosen reservoir $\phi_s^{\text{res}}$. In practice, we start by calculating $\mu_s$ for the desired values of $\phi_s^{\text{res}}$ and for a few values of $\phi_s$ for the value of $\phi_l$ we are considering. The results are interpolated with cubic splines, and an approximation to $\phi_s$ is read by running a horizontal line at the previously calculated value of $\mu_s$. A new calculation of $\mu_s$ is carried out at that value of $\phi_s$, and the point so obtained is added to the accumulated list of $\phi_s, \mu_s$, and a new cubic-spline fitting is applied. The procedure is repeated until we have points in the list mentioned above that are very close in the vertical direction to the assigned value of $\mu$. 

Once we obtain the appropriate values $\phi_s(\phi_l)$, we run the Molecular Dynamics in the given triplet of parameters $q$, $\phi_l$, and $\phi_s(\phi_l, \phi_s^{\text{res}})$. There is some small error introduced by the fact that, being integers $N_s$ and $N_l$, we generally cannot generate simultaneously the correct values of $\phi_s$ and $\phi_l$. In this case, we have given priority to $\phi_l$, making it exact to numerical precision, and absorbed the small error in $\phi_s$, given the fact that $N_s$ is usually much larger than $N_l$ and less affected by a $\Delta N = 1$ error.

For each triplet of parameters (one $q$ and two packing fractions) that we have explored, we have run ten realizations of the simulations, both in the calculations of $\mu_s$ and in the actual measurement of forces, to improve statistics and gain an understanding of error levels. All simulations have run on GPUs, using the CUDA extension of the C language. In all cases, we have used a minimum of 2,000 large particles (sometimes many more), which sometimes implies that we might have to employ 200,000 or more small spheres.

\subsection{Chemical potentials and the calculation of $\phi_s(\phi_l,\phi_s^{\text{res}})$}

\begin{figure}
\includegraphics[width = 0.47\textwidth]{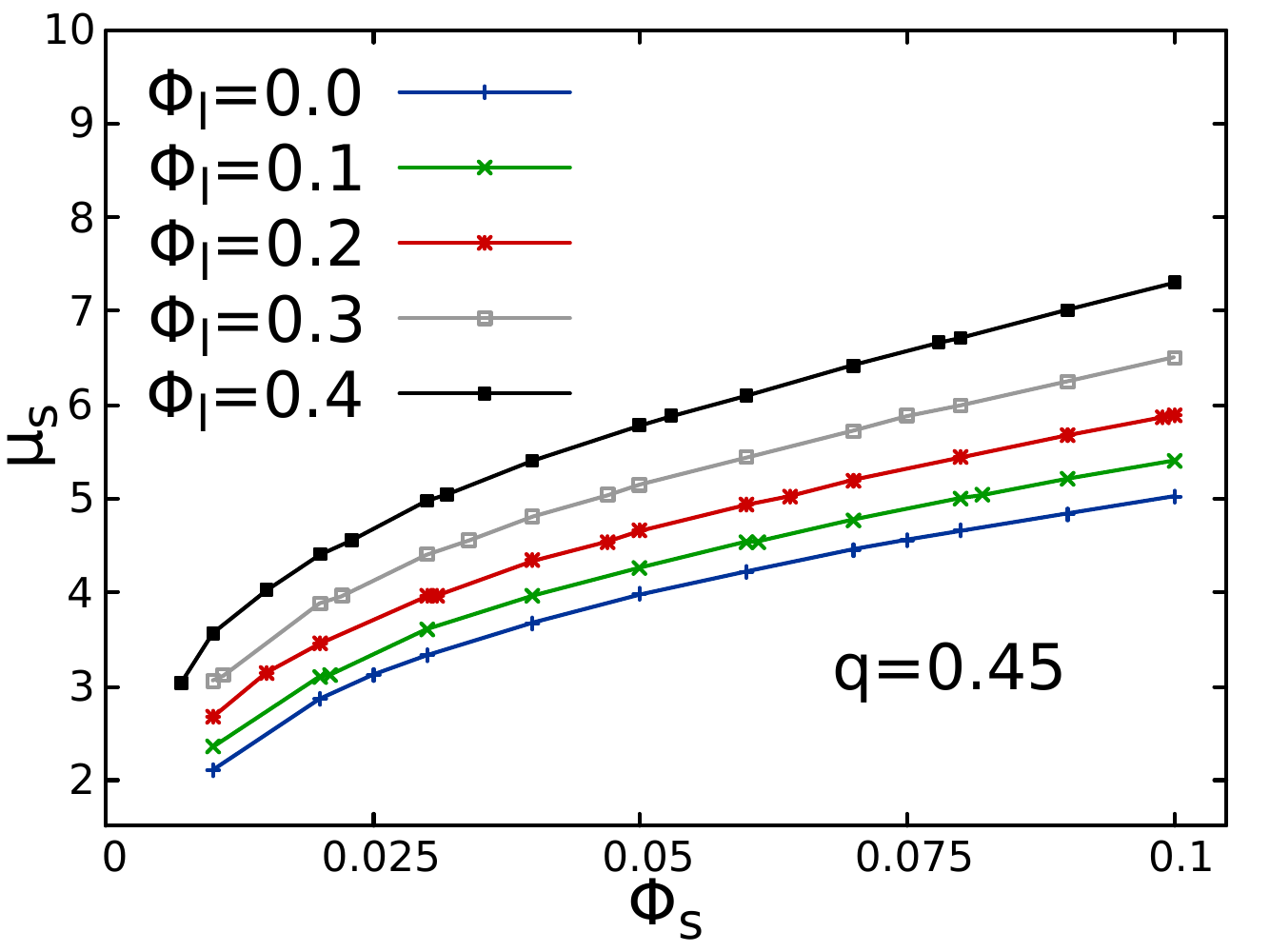}
\caption{Curves of $\mu_s$ vs. $\phi_s^{\text{res}}$ for
	different $\phi_l$, as denoted, for the size asymmetry $q=0.45$. 
     Error bars are much smaller than the symbols. The lines are cubic splines, and were used in the construction of the $\phi_s(\phi_l)$ \emph{vs.\/} $\phi_s^{\text{res}}$ correspondence.}
\label{fig1}
\end{figure}

\begin{figure}
\includegraphics[width = 0.47\textwidth]{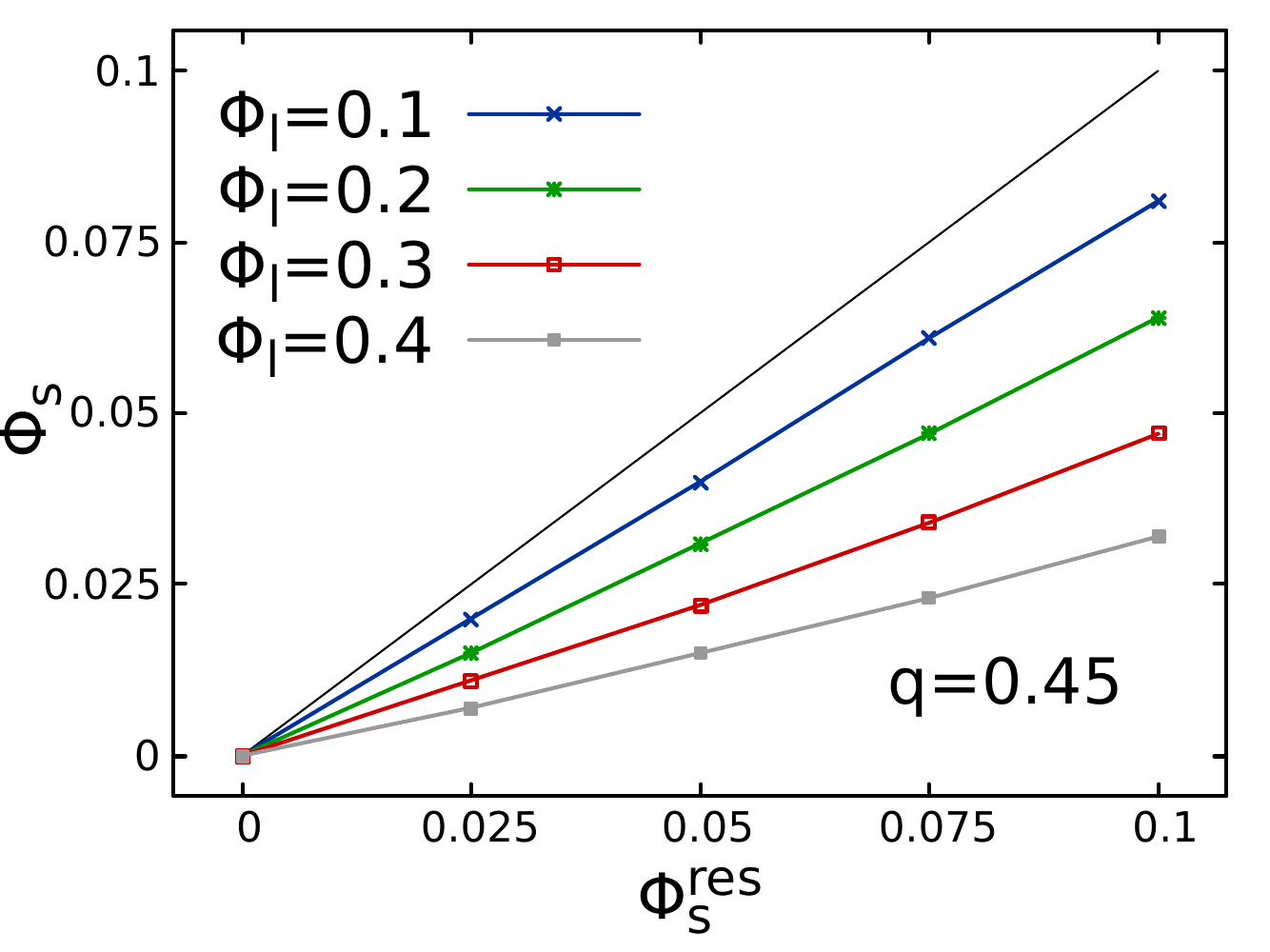}
	\caption{Curves of $\phi_s$ vs. $\phi_s^{\text{res}}$ for several $\phi_l$, as indicated, for the size asymmetry $q=0.45$.  
    The lines are cubic splines, given as a guide to the eye. The thin black line has slope 1 and gives the trivial case $\phi_l = 0$.}
\label{fig2}
\end{figure}

We have worked in the semi-grand ensemble, where, as stated before, in the limit of small depletants ($q < q_3$), one finds that the depletion forces become independent of the density of the depleted particles, as long as the chemical potential $\mu_s$ for the depletants remains constant. The packing fractions $\phi_s$ were adjusted as previously explained to maintain the constant $\mu_s$ condition; examples of the behavior of $\mu_s$ \emph{vs.\/} $\phi_s^{\text{res}}$ and $\phi_s(\phi_l)$ \emph{vs.\/} $\phi_s^{\text{res}}$ for the same $\mu_s$ are given in Figs.~\ref{fig1} and~\ref{fig2}, respectively, for the case $q = 0.45$. The values obtained for all combinations of parameters are given in Table~\ref{phi-sml}.

\begin{table*}
\caption{\label{phi-sml}
	Values of $\phi_s$ calculated for a given pair $\phi_s^{\text{res}}, \phi_l$. The last row of each group gives the $\mu_s$ values.}
\begin{ruledtabular}
\begin{tabular}{|c|cccc|}
        & & $q=0.15$ & & \\
        \hline
        \diagbox{$\phi_l$}{$\phi_s^{\rm res}$} & 0.025 & 0.050 & 0.075 & 0.100 \\
        \hline
        0.1 & 0.02223 & 0.04460 & 0.06707 & 0.08965 \\
        0.2 & 0.01949 & 0.03924 & 0.05921 & 0.07936 \\
        0.3 & 0.01676 & 0.03392 & 0.05137 & 0.06916 \\
        0.4 & 0.01408 & 0.02867 & 0.04364 & 0.05907 \\
        \hline
        $\mu_s$ & 11.341 & 12.199 & 12.783 & 13.262 \\
        \hline \hline
        & & $q=0.45$ & & \\
        \hline
        0.1 & 0.01973 & 0.03992 & 0.06052 & 0.08148 \\
        0.2 & 0.01480 & 0.03035 & 0.04666 & 0.06363 \\
        0.3 & 0.01036 & 0.02158 & 0.03373 & 0.04682 \\
        0.4 & 0.00668 & 0.01408 & 0.02238 & 0.03158 \\
        \hline
        $\mu_s$ & 3.1011 & 3.9597 & 4.5430 & 5.0220 \\
        \hline \hline
        & & $q=0.60$ & & \\
        \hline
        0.1 & 0.01816 & 0.03697 & 0.05639 & 0.07633 \\
        0.2 & 0.01219 & 0.02530 & 0.03935 & 0.05436 \\
        0.3 & 0.00727 & 0.01567 & 0.02485 & 0.03513 \\
        0.4 & 0.00371 & 0.00822 & 0.01355 & 0.01976 \\
        \hline
        $\mu_s$ & 0.94353 & 1.8021 & 2.3854 & 2.8643 \\
\end{tabular}
\end{ruledtabular}
\end{table*}

\section{Effective two-body depletion forces: the role of many-body contributions}

\subsection{\label{cb} Distinguishing higher-order contributions in the effective two-body force}

Appendix \ref{DF_CBF} provides a concise overview and introduces the essential components necessary to characterize effective interactions within the contraction of bare forces (\ref{DF_CBF}.1) \cite{delossantos-lopez_JCP_2022} and explores the framework of integral equation theory (\ref{DF_CBF}.2) \cite{alcaraz-klein}. This framework has recently been applied to enhance our understanding of the influence that particle concentration has on depletion forces \cite{delossantos-lopez-PA-2024}. Equations (\ref{dp1}) and (\ref{ceff}) indicate that depletion potentials are expressed through direct correlation functions, $c_{ij}(r)$, and bridge functions, $b_{ij}(r)$ (for further explanation, see Appendix \ref{DF_CBF}). Additionally, a brief and superficial examination of the diagrammatic representations of these functions,\cite{Hansen1986}
\begin{equation*}
   c(r_{12}) = \includegraphics[scale=0.1]{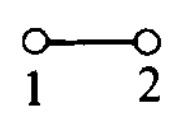} + \includegraphics[scale=0.1]{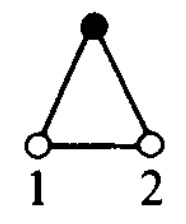} + \cdots
\end{equation*}
and
\begin{equation*}
    b(r_{12})  =  \includegraphics[scale=0.1]{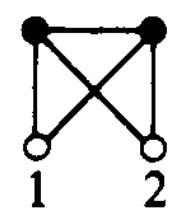} + \cdots,
\end{equation*}
reveals that the leading contributions to the direct correlation functions come from pair and triplet interactions, while the primary term in the bridge functions pertains to quadruplet configurations. Consequently, omitting the bridge functions from Eq. (\ref{dp1}) will, at least in part, reduce the influence of such higher-order configurations on the effective interactions evaluated. This hypothesis will be examined in another contribution.

\subsection{Depletion forces intrinsically including many-body contributions}

From now on, we will display the results for the depletion forces. However, before we do so, we need to address a pertinent question: Why not simply perform a numerical integration of these forces and subsequently display the resulting effective potentials? This would allow for a direct comparison with results reported elsewhere. In fact, numerical integration of noisy data has the nice characteristic of reducing the level of noise. However, the negative side of this feature is that, in general, this procedure tends to develop a drift, caused by the slow accumulation of the integration of errors, providing a random walk-like addition to the desired signal\cite{Kloeden1992}. There is also the obvious matter of an unknown global additive constant. Since we are primarily concerned with the possible separation of forces (and, by extension, potentials), the presence of these uncontrolled drifts would render the results for potentials completely inconclusive.

\begin{figure}
\includegraphics[width = 0.45\textwidth]{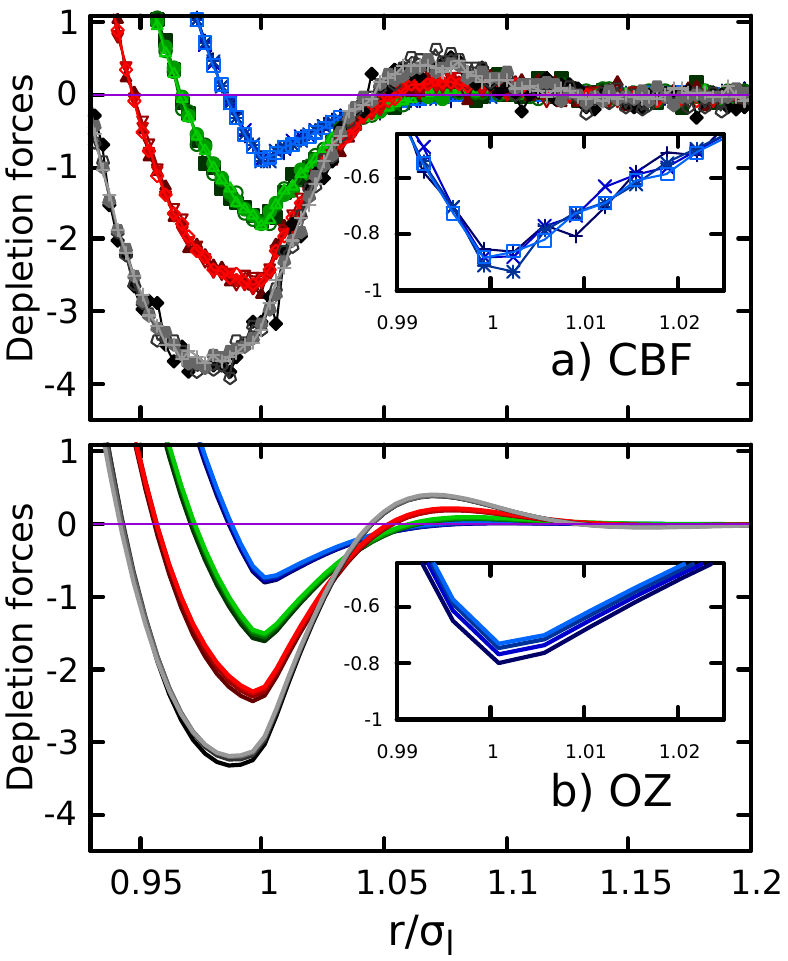}
	\caption{Depletion forces, $f_d$, between large colloids immersed in a bath of small ones for a size asymmetry $q = 0.15$, obtained from the (a) CBF approach and the (b) integral equations formalism for effective interactions, with $\phi_s^{\text{res}} = 0.025$, $0.05$, $0.075$ and $0.1$ which correspond to the blue, green, red and gray lines, respectively. Each group of curves represents a superposition of results for $\phi_l = 0.1$, $0.2$, $0.3$, and $0.4$, which, on the color scale, transition from light to dark; the results become indistinguishable within the level of numerical error. Insets show an enlargement of the negative-forces region for $\phi_s^{\text{res}} = 0.1$; it is clear that there is no systematic ordering of the four different lines.} 
\label{fig3}
\end{figure}

Let us start with the results for $q = 0.15$, as shown in Fig.~\ref{fig3}. Let us first pay attention to the simulation results obtained from the CBF approach, see Fig.~\ref{fig3}(a). With the caveat that the lines for the depletion forces $f_d$ remain noisy, we do not find any evidence of a $\phi_l$ dependence, even if the depletant particles do not behave as an ideal gas and are instead close to hard spheres. The depletion force curves for all values of $\phi_l$ superpose up to numerical error, implying that the higher-order diagrams with three large spheres and two or more small ones, as described in \cite{dijkstra_PRE_1999}, are of no importance. As the density of depletants in the reservoir increases, we find that the main minimum (the most attractive force) pushes down, and we also find the development of a small hump in the force, that is, a weak repulsive region; these two changes are well-known effects of the increase in depletants (in this case, $\phi_s^{\text{res}}$). The effects of the softness of the potential are clear in two points: the finite value slopes of the repulsive forces at the left extreme,  and the increasing smoothing of curves for the attracting forces, which imply a corresponding smoothing of the potentials. For the smallest $\phi_s^{\text{res}}$, we can still see the inverted cusp corresponding to the contact between large spheres (see the inset of Fig.~\ref{fig3}(a). The width of the negative force region is around $0.11\sigma_l$, generally smaller than the diameter of the depletants, which for hard spheres should be close to the width of the attractive zone. The softness of the potential has reduced this distance. Lastly, note that we only show the $r$ values in the vicinity of the $\sigma_l$ region (where negative forces are present). To the left of this region, one basically gets the original repulsive mWCA force, and to the right, the forces are null.

Fig.~\ref{fig3}(b) displays the depletion forces obtained from the integral equations formalism for the cases reported in Fig.~\ref{fig3}(a). Interestingly, one can notice the high level of accuracy of the integral equations theory for effective interactions, which quantitatively reproduces the simulation results. We should emphasize that the theoretical formalism is derived in the canonical ensemble, which means that the inputs needed to evaluate the depletion potential, $\beta u^{\mathrm{eff}}_{ll}(r)$, and from it the depletion force, are essentially the bare potential between particles (Eq. (\ref{mWCA})), and the packing fractions of the large and small particles, as reported in Table \ref{phi-sml}. This excellent agreement between simulation and theory highlights the accuracy of the mVerlet closure (Eq. (\ref{BmV})) even for soft repulsive potentials.

\begin{figure}
\includegraphics[width = 0.45\textwidth]{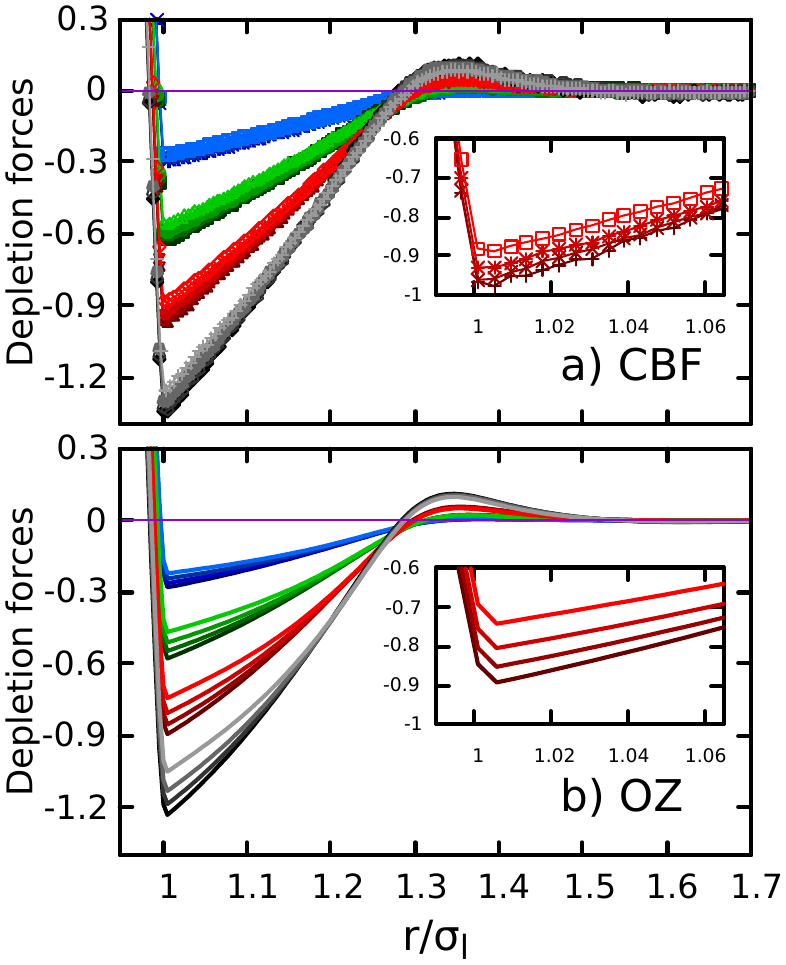}
	\caption{Depletion forces, $f_d$, between large colloids immersed in a bath of small ones for a size asymmetry $q = 0.45$, obtained from the (a) CBF approach and the (b) integral equations formalism with $\phi_s^{\text{res}} = 0.025$, $0.05$, $0.075$ and $0.1$ which correspond to the blue, green, red and gray lines, respectively. Each group of curves corresponds to the results for $\phi_l = 0.1$, $0.2$, $0.3$, and $0.4$, that on the color scale go from light to dark, 	For each $\phi_s^{\text{res}}$ the lines for different $\phi_l$ separate, although remaining close to each other. This separation is roughly proportional to the magnitude of the force. However, the crossing point of these lines does not coincide with zero force, and happens instead at small negative values of $f_d$ (see inset in Fig.~\ref{fig5}(a). Insets show an enlargement of the negative-forces region for $\phi_s^{\text{res}} = 0.075$.}
\label{fig4}
\end{figure}

\begin{figure}
\includegraphics[width = 0.45\textwidth]{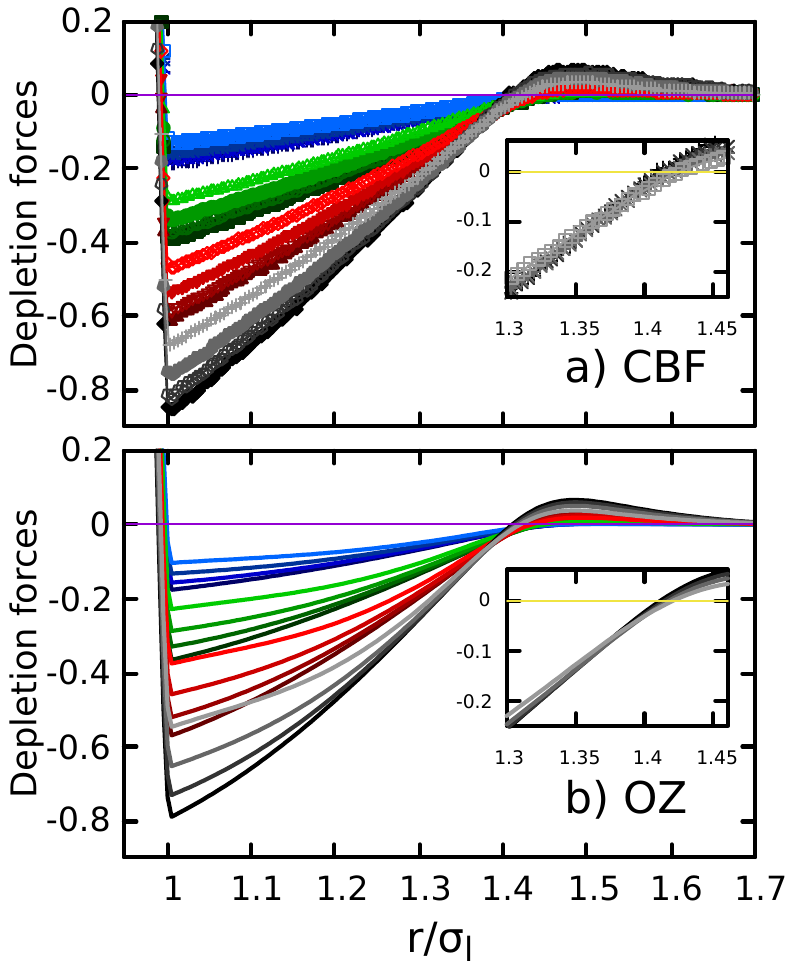}
	\caption{Depletion forces, $f_d$, between large colloids immersed in a bath of small ones for a size asymmetry $q = 0.6$, obtained from the (a) CBF approach and the (b) integral equations formalism. Values of $\phi_s^{\text{res}}$ and $\phi_l$ are as in Fig. \ref{fig4}. For each $\phi_s^{\text{res}}$, the lines for different $\phi_l$ are now widely separated. This separation is again roughly proportional to the magnitude of the force, but the crossing point of these lines does not coincide with zero force, and happens instead at small negative values of the depletion forces, as can be seen in the insets, which show the crossing region for the forces for the $\phi_s^{\text{res}} = 0.1$ case.}
\label{fig5}
\end{figure}

Now we discuss the results for the two other values of $q$ covered, namely $q = 0.45$ and $0.6$. For these cases, we get very clear effects of the large particle density on the forces, which are displayed in Figs.~\ref{fig4} and~\ref{fig5}, respectively. The general features of the forces are similar to those of the $q = 0.15$ case, that is, an attractive interval (negative forces) of a width smaller than the diameter of the depletants, and a repulsive hump (positive forces). Since the overlaps between large spheres are now much smaller than the diameters of small ones, the inverted cusps corresponding to the large-large contact remain very sharp. For all values of $\phi_s^{\text{res}}$, the magnitude of the forces decreases as the value of $\phi_l$ increases; this is a feature that has already been observed in previous works \cite{goulding_PRE_2001,nguyen_SM_2018,lowen_JP:CM_1998}. 

The insets in Figs.~\ref{fig4}(a) and~\ref{fig4}(b) show an enlargement of the force for $\phi_s^{\text{res}} = 0.075$. The lower curve for $\phi_l = 0.1$, obtained from the CBF approach, is still very noisy, while the other curves become cleaner as $\phi_l$ increases. This happens simply because within the CBF approach, the number of large colloids is always $N_l > 2000$, and gathering enough statistics becomes easier for larger values of $\phi_l$. However, simple observation shows that the neighboring curves are not equidistant, but separate more from each other as $\phi_l$ grows, as expected from the proposed polynomial dependence on $\phi_l$ (Eq.~(\ref{force-polynomial})), and also captured (and more notorious) by the predictions from the integral equations framework. 

The simulation results obtained for the binary mixture with $q = 0.6$ show more dramatically the effects of $\phi_l$ on the depletion forces, see Fig.~\ref{fig5}(a). We observe a clear separation of the different curves, indicating a strong dependence on the density of large particles. This is also confirmed by the integral equations approximation for the effective interactions, see Fig.~\ref{fig5}(b). As before, the separation between lines increases with the value of $\phi_l$, at the same time that the magnitude of the depletion force decreases. It becomes much easier to notice that the crossing region of the curves for the depletion forces is not around $f_d = 0$, as shown in the insets of the figures. This means that a simple re-scaling of the forces
cannot be completely successful.

\begin{figure}
\includegraphics[width = 0.50\textwidth]{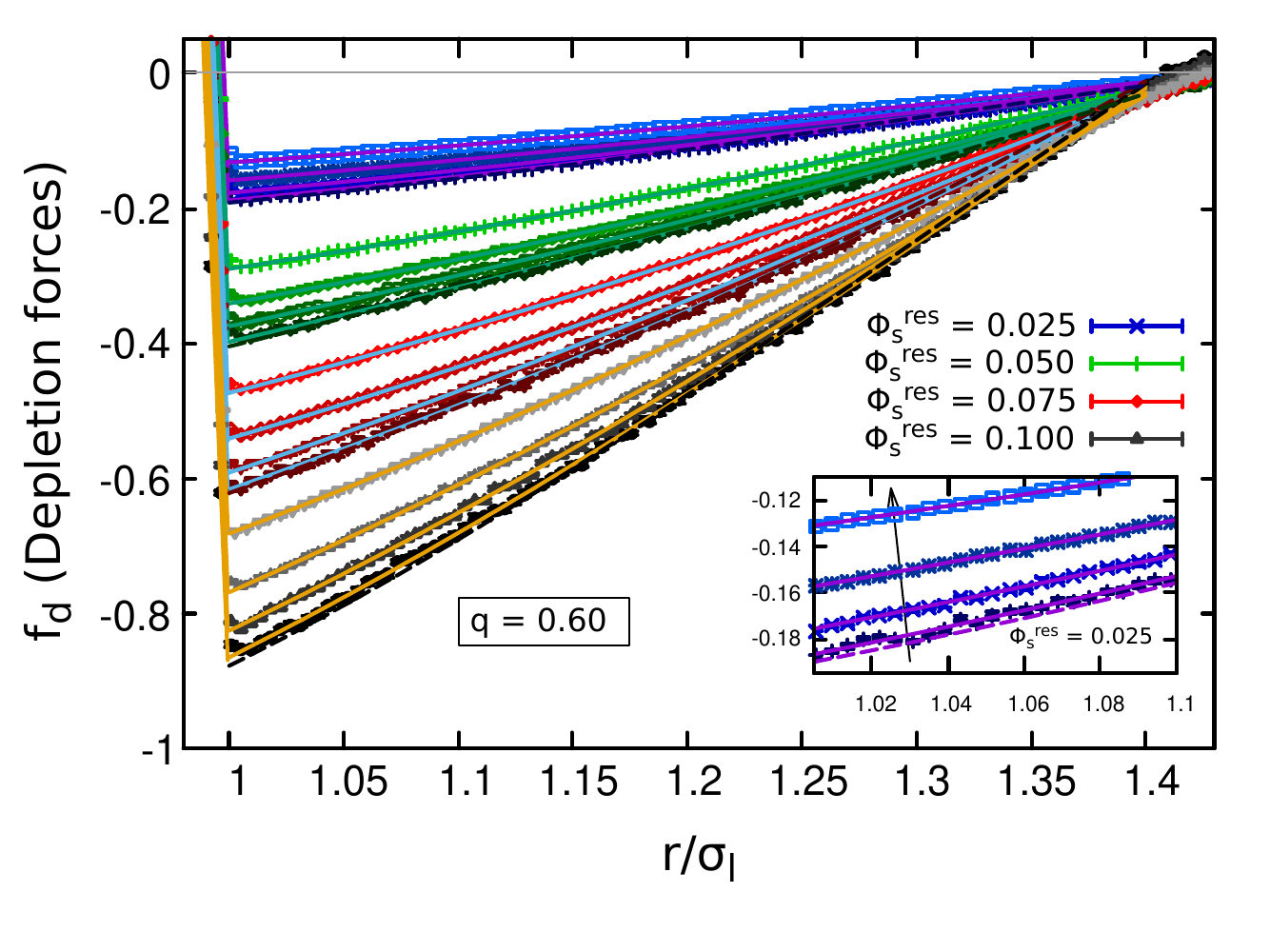}
	\caption{Curves of the attractive part of the depletion forces for	$q = 0.6$ obtained from the CBF approach. Values of $\phi_s^{\text{res}}$ and $\phi_l$ are as in previous figures. For each value of $\phi_s^{\text{res}}$, the lines for different $\phi_l$ have been fitted to a third-order degree polynomial, for the $1.0 < r/\sigma_{l} < 1.4$ interval. The coefficients of the different fits have been used to extrapolate to $\phi_l = 0$, and the results are given by the dashed lines just below the line corresponding to $\phi_l = 0.1$ (bottom line in each group of four). Insets allow for a better view of the four depletion forces, together with the extrapolated one, for $\phi_s^{\text{res}} = 0.025$.}
\label{fig6}
\end{figure}

Both cases $q = 0.45$ and $q = 0.6$ show an attractive almost linear ramp in the region where the forces are stronger, looking a bit similar to the AO limit \cite{asakura_JPS_1958,alcaraz-klein,delossantos-lopez-PA-2024}. Now, we can take advantage of that quasi-linear behavior to do a simple, low-degree polynomial fitting of the curves, at least by sections, since no simple polynomial fitting can handle the cusp in $r/\sigma_{l} \sim 1$. The results are shown in Fig.~\ref{fig6}, for third-order polynomials covering the $1 < r/\sigma_{l} < 1.4$ interval; and the coefficients are given in Table \ref{tableofcoefficients} (see Appendix \ref{polynomial_fit}). Now, an advantage of these fittings is that they give us a collection of coefficients that will depend on the packing fraction of large spheres, $\phi_l$. A second fitting, now of the coefficients themselves as functions of $\phi_l$, can be used to extrapolate to the limit $\phi_l \to 0$, allowing us to predict in this way the shape of the depletion forces-curve for the infinitely diluted limit. We have performed this extrapolation for the negative ramp in the forces for $q = 0.6$; the results are given in the dashed lines of Fig.~\ref{fig6}. The distances between these extrapolated lines, for $\phi_l \to 0$, and those for $\phi_l = 0.1$, are tiny; a better view of the four depletion forces and one extrapolated, for the case $\phi_s^{\text{res}} = 0.025$, is given in the insets of Fig.~\ref{fig6}.

For cases $q=0.45$ and $q=0.60$, the integral equation formalism for effective interactions also quantitatively reproduced the physical scenario reported by the CBF approach, although in these cases, many-body contributions become much more evident than in the case $q=0.15$. This means that although the OZ equation describes two-body correlations, those functions contain higher-order contributions mediated by the pair potential and are captured through the effective two-body force between large particles, an aspect that has not been discussed before \cite{ramon-ro-al-2,delossantos-lopez_JCP_2021,delossantos-lopez_JCP_2022,delossantos-lopez-PA-2024}.

\subsection{Depletion forces without including many-body interactions}

\begin{figure}
    \centering
    \includegraphics[width=0.7\linewidth]{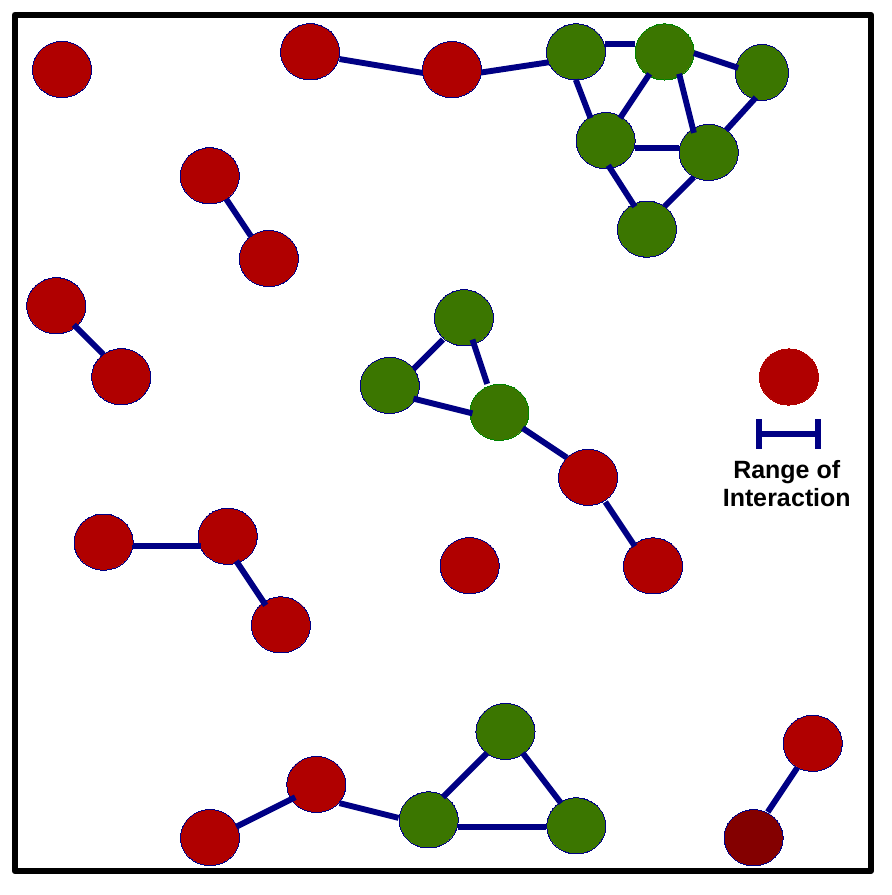}
    \caption{Representative diagram of how we proceed in computational simulations to eliminate many-body interactions for the calculation of the depletion forces between large particles. The sets of particles that form triangles (particles in green) are discarded when considering the interactions in Eq. (\ref{ftot}). Only the particles whose configuration is that of the particles in red are taken into account when calculating the total force, see, e.g., Appendix \ref{DF_CBF}.1.}
    \label{fig7}
\end{figure}

To quantify the role of many-body interactions in the depletion force, we compare results obtained with and without their inclusion. While previous sections emphasized the relevance of such interactions, here we aim to isolate their effects by removing them from the calculation. In the CBF scheme, this is achieved by modifying the sampling criteria: Only configurations where large particles do not form triangles are included (structures of higher order are also excluded due to the presence of triangular substructures within them); the algorithm is explicitly discussed in Fig.~\ref{fig7}.

In contrast, the OZ theoretical framework inherently includes many-body interactions through both the direct correlation functions and the bridge functions, as discussed previously. Since no systematic procedure currently exists to construct closure relations that explicitly exclude many-body contributions, we do not attempt to modify the closure relation in this work. However, since the bridge function is made up entirely of diagrams that involve triangular loops \cite{Hansen1986} (actually, the leading term refers to the configurations of four particles), we remove it from Eq.~(\ref{dp1}) to partially eliminate many-body effects from the effective force (according to the qualitative analysis presented in Subsection \ref{cb}). In other words, the influence of many bodies persists, albeit to a diminished degree.

\begin{figure}
\includegraphics[width = 0.45\textwidth]{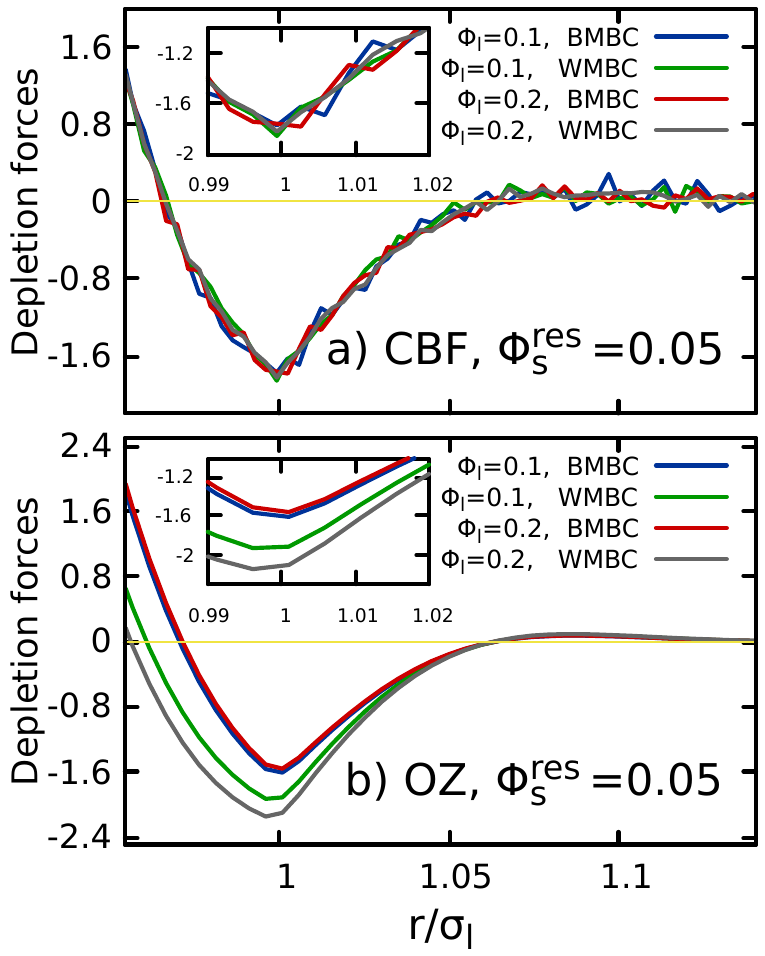}
	\caption{Depletion forces, $f_d$, between large colloids immersed in a bath of small ones for a size asymmetry $q = 0.15$, obtained from the (a) CBF and the (b) OZ equation. Values of $\phi^{\text{res}}_s$ and $\phi_l$ are indicated. The blue and red curves correspond to the calculation of the forces considering the contributions of many bodies, while in the green and gray curves, the interactions of many bodies are discarded. MBC stands for ``by many-body contributions", while noMBC stands for ``without many-body contributions".}
\label{fig8}
\end{figure}

Fig. \ref{fig8}(a) shows the depletion force obtained using the CBF scheme for $q=0.15$ and $\phi_l = 0.1, 0.2$, comparing the models with many-body contributions (MBC) and without many-body contributions (noMBC). Only small differences are observed between the two models at both concentrations. This is consistent with previous theoretical predictions suggesting that for $q \leq 0.15$, the contribution of triangular diagrams is negligible \cite{dijkstra_PRE_1999}. The remaining differences may arise from concentration effects rather than genuine many-body interactions.

In Fig.~\ref{fig8}(b), the MBC and noMBC models are compared using the OZ approach for the same parameters $q$ and $\phi_l$. Although the MBC model exhibits similar behavior for both values of $\phi_l$, a noticeable discrepancy appears between the MBC and the noMBC models. Specifically, results show a reduction in the attractive force when many-body interactions are omitted. This contrasts with the CBF results, which highlight a limitation of our approach: Eliminating the bridge function is insufficient to completely remove the many-body effects, as their impact continues to be reflected in the outcomes.

\begin{figure}
\includegraphics[width = 0.45\textwidth]{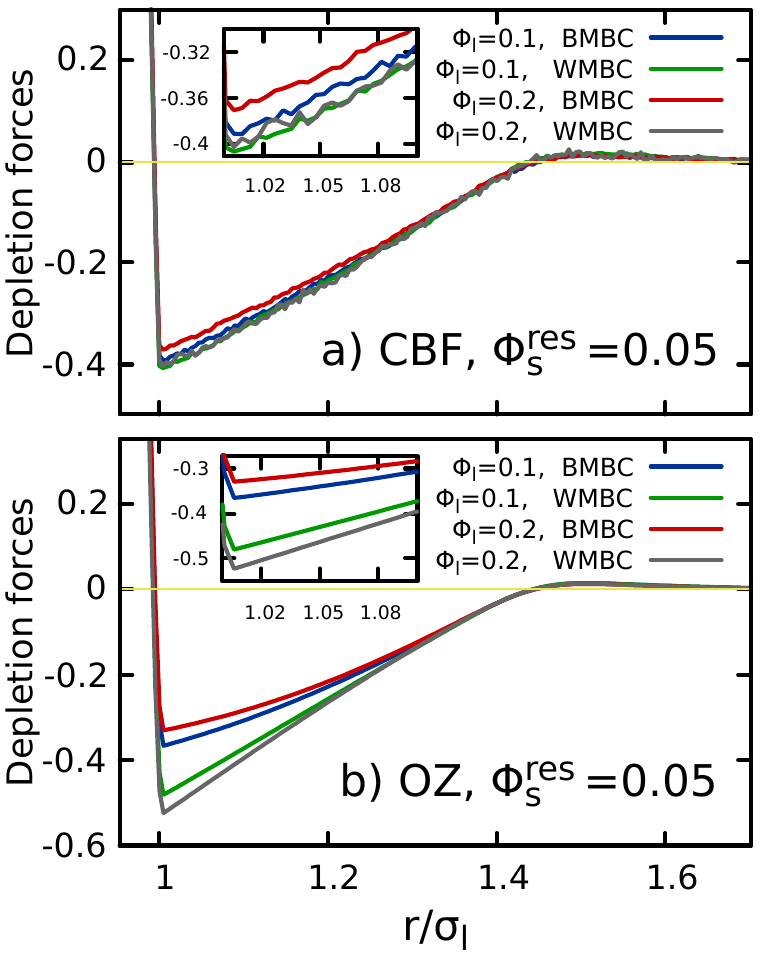}
	\caption{Depletion forces, $f_d$, between large colloids immersed in a bath of small ones for a size asymmetry $q = 0.60$, obtained from the (a) CBF and the (b) OZ equation. Values of $\phi^{\text{res}}_s$ and $\phi_l$ are indicated. The blue and red curves correspond to the calculation of the forces considering the contributions of many bodies, while in the green and gray curves, the interactions of many bodies are discarded. MBC stands for ``by many-body contributions", while noMBC stands for ``without many-body contributions".}
\label{fig9}
\end{figure}

For $q = 0.60$, Fig.~\ref{fig9}(a) presents the depletion force calculated using the CBF approach. The noMBC model shows virtually identical results for both concentrations, within numerical error. However, in the MBC model, the force becomes less attractive as $\phi_l$ increases, clearly deviating from the behavior of noMBC. This reflects the growing influence of many-body interactions at higher size ratios and concentrations.

The corresponding results of the OZ scheme for $q = 0.60$ are shown in Fig.~\ref{fig9}(b). As in $q = 0.15$, the MBC force is systematically less attractive than the no MBC force. Compared to the CBF results, the magnitude of the deviation suggests an even stronger many-body contribution. However, for the aforementioned reasons, it is essential to exercise caution when interpreting predictions based on OZ in this context.

\section{Conclusions and outlook}
We have successfully quantified the many-body contributions to the depletion forces in a binary mixture of quasi-hard spheres. We have systematically explored several values of the packing fractions for large and small particles ($\phi_l$ and $\phi_s$, respectively) and have verified, with numerical precision, that for size ratios $q = \sigma_s/\sigma_l \le q_3 = 2/\sqrt{3} -1 \approx 0.1547$ the depletion forces become independent of $\phi_l$. This means, therefore, that no many-body terms are necessary to describe the interaction in this situation, even if the theoretical approaches indicate that for hard-sphere depletants there may be some ---perhaps very weak --- effect of three-body and higher order interactions. 

However, the results obtained for $q = 0.45$ and $q = 0.6$ show a strong dependence on the population of large particles, introducing, for $q = 0.6$, a spread of approximately 20\% in the forces for a change from $\phi_l = 0.1$ to $\phi_l = 0.4$. Moreover, we find that these changes become much more pronounced as $\phi_l$ increases, supporting the proposed polynomial dependence on $\phi_l$. As confirmed in several previous works, the strength of the depletion forces decays with $\phi_l$, indicating some degree of screening of large particles in comparison to those considered in a binary interaction. The behavior with respect to the reservoir concentration $\phi_s^{\text{res}}$ is close to what is expected from an AO model: the force strength is almost --- but not perfectly --- proportional to the depletant population in the reservoir.

A peculiar question appears when we take into account that for $q > q_3$ larger densities of depleted particles imply a weakening of the depletion attraction: Any possible phase change that depends on the depletion attraction over the large particles is going to be frustrated by the fact that denser agglomerations of them will reduce their attraction. It may be possible then that around the transition values of $\phi_l$ and $\phi_s$ for a fluid-fluid, or even a fluid-solid transition, the fluctuations become quite large, since condensation implies weaker attraction, while dilution gives rise to stronger attraction. This is a phenomenon that may warrant extensive exploration.

\begin{acknowledgments}
The authors thank the financial support from SECIHTI-Mexico (Grant No. CBF2023-2024-3350). N.M.S.-L. also thanks SECIHTI for financial support (postdoctoral scholarship 2025).
\end{acknowledgments}

\appendix

\section{Determining Depletion Forces in Binary Mixtures}\label{DF_CBF}
\subsection{Contraction of the bare forces}

Effective interactions, and in particular depletion interactions, can be determined at the level of forces using an average of the effects of the small particle-big particle forces in the dynamics of the large ones. This is an approach that we have named \emph{contraction of the forces\/} (CBF) in previous works~\cite{delossantos-lopez_JCP_2021,delossantos-lopez_JCP_2022,pedrozo-romero_JCP_2024,delossantos-lopez-PA-2024}. For completeness and for the sake of discussion, we will briefly review the approach here.

To implement the contraction of the forces, assume a colloidal mixture made up of $N_l$ large spheres and $N_s$ small ones. Also, consider that the \emph{total\/} force exerted on the $i$-th large particle, ${\bf F}^l_i ({\bf r}_i )$, at some time $t$ can be measured. This is straightforward for simulations and, in principle, possible in colloidal experiments by carefully tracking the particle trajectory and evaluating its acceleration from second differences in position. The results will generally be quite noisy, but a reasonable signal may be extracted from the averages. Now, this force is given by
\begin{equation}
	{\bf F}^l_i ({\bf r}_i ) = \sum_{j \neq i}^{N_l} {\bf f}^{ll}_{ij}({\bf r}_{ij}) +
	\sum_{k=1}^{N_s} {\bf f}^{sl}_{ik}({\bf r}_{ik}).
	\label{ftot}
\end{equation}
As usual, ${\bf f}^{ll}_{ij}({\bf r}_{ji})$ is the bare force exerted by the $j$-th large particle and ${\bf f}^{sl}_{ik}({\bf r}_{ik})$ the one exerted by the $k$-th small particle, all of them over the $i$-th particle. Now, in the contracted description, the previous equation (\ref{ftot}) is written as,
\begin{equation}
	\langle {\bf F}^l_i ({\bf r}_i ) \rangle_s = 
	\sum_{j \neq i}^{N_l} {\bf G}^{ll}_{ij}({\bf r}_{ij}),
	\label{ftotcont}
\end{equation}
where $\langle {\bf F}^l_i ({\bf r}_i ) \rangle_s$ is now the average of ${\bf F}^l_i ({\bf r}_i )$ in the configurations of the small particles, for any ensemble that one may be using, in the fixed field of the large particles. Therefore, $u^{\text{eff}}_{ll}(r_{ij})$ in ${\bf G}^{ll}_{ij}({\bf r}_{ij}) = G^{ll}_{ij}(r_{ij}) \hat{\bf r}_{ij}=- \bm{\nabla} u^{\text{eff}}_{ll}(r_{ij})$ corresponds to the theoretical effective interaction potential between large particles. $r_{ij}$ is the magnitude of ${\bf r}_{ij}$, $\hat{\bf r}_{ij} = {\bf r}_{ij}/r_{ij}$, and $\bm{\nabla} = d/d{\bf r}_{ij}=\hat{\bf r}_{ij} d/dr_{ij}$.

Notice that in the last term of the previous equation we are making two fundamental assertions: (a) given the isotropy of space, the components of ${\bf G}_{ij}$ that are normal to the unitary vector ${\bf r}_{ij}$ should cancel out and at the end, taking into account the averaging process, ${\bf G}_{ij} = G_{ij} {\bf \hat r}_{ij}$. (b) We also declare that the effective force $G_{ij}$ will depend only on the distance $r_{ij}$, as expected from the very definition of the effective force.

The small particles are traced out from the dynamics by going from Eqs.~(\ref{ftot}) to~(\ref{ftotcont}), and so, after averaging over their configurations, they become part of the supporting environment instead of being an explicit part of the colloidal mixture. This environment changes as the positions of the large particles change, and its effects are implicitly included in the effective force ${\bf G}^{ll}_{ij}({\bf r}_{ij})$ between them. The idea is to evaluate ${\bf F}^l_i ({\bf r}_i )$ using Eq.~(\ref{ftot}), and then to take it into Eq.~(\ref{ftotcont}) to get ${\bf G}^{ll}_{ij}({\bf r}_{ij})$ through a ``best fit'' process. The instantaneous values of ${\bf F}^l_i ({\bf r}_i)$ obtained from Eq.~(\ref{ftot}) are noisy, even if they are obtained through a deterministic simulation run, and only their averages over long times will satisfy Eq.~(\ref{ftotcont}). They may then be described by the stochastic equation,
\begin{eqnarray}
    {\bf F}^l_i ({\bf r}_i ) &=& \sum_{j \neq i}^{N_l} {\bf G}^{ll}_{ij}({\bf r}_{ij})+{\bf D}^{l}_{i}(t) \nonumber \\ &=&
	\sum_{j \neq i}^{N_l} G^{ll}_{ij}(r_{ij}) {\bf \hat r}_{ij} +{\bf D}^{l}_{i}(t),
	\label{ftotcont2}
\end{eqnarray}
where ${\bf D}^{l}_{i}(t)$ is the non-systematic part of the instant forces, that is, the part that is due to the constant tapping of small particles on the large ones. This is expected to be a noisy term with zero mean $\langle {\bf D}^l_i (t) \rangle_s = 0$, and finite variance (although this last point is irrelevant for the determination of the effective interaction). 

With the given decomposition, the force ${\bf F}^l_i ({\bf r}_i )$ in Eq.~(\ref{ftotcont2}) can be interpreted as the noisy result of a measurement and the $G^{ll}_{ij}(r_{ij})$'s as fitting functions for it, with ${\bf D}^{l}_{i}(t)$ being the instant deviations between the measured and fitted values. Therefore, the values of $G^{ll}_{ij}(r_{ij})$ can be obtained using the least-squares (LS) method~\cite{press_1992}. We would like to make clear that the functions over which the LS expansion is carried out are not continuous; in fact, we are using impulses distributed in the interparticle distance covered by the force.

We will now concentrate on the study of depletion forces in binary mixtures. The application of the contraction of forces given by Eq.~(\ref{ftotcont2}) proceeds by getting ${\bf F}^l_i ({\bf r}_i )$ from Eq.~(\ref{ftot}) and splitting Eq.~(\ref{ftotcont2}) into its Cartesian components,
\begin{equation}
	F_{i,\alpha}^l ({\bf r}_i ) = 
	\sum_{j\neq i}^{N_l} G^{ll}_{ij} (r_{ij}) \cos \theta_{ij,\alpha} + D_{i,\alpha}^l (t),
	\label{fsplited}
\end{equation}
with $\alpha=x,y,z$ and $\cos \theta_{ij,\alpha}$ the corresponding direction cosines. To proceed numerically, we discretize the distance $r_{ij}$ into $C$ classes,
\begin{equation}
	r_{ij} \rightarrow r_c = c \delta, \quad \mbox{if} \quad r_c \le r_{ij} < r_{c+1},
	\label{clases}
\end{equation}
where $\delta$ is the class size and $c = 0,...,C-1$, and it is assumed that $r_{max} = C \delta$ is the range of the depletion effects. Within this discretization, the forces become,
\begin{equation}
	F_{i,\alpha}^l ({\bf r}_i ) = \sum_{c=0}^{C-1} A_{c,\alpha}^i G^{ll}_{ij} (r_c) + D_{i,\alpha}^l (t),
	\label{fsplited2}
\end{equation}
with
\begin{equation}
	A_{c,\alpha}^i = \sum_{j\neq i}^{N_{l,c}} \cos \theta_{ij,\alpha}.
	\label{fsplited3}
\end{equation}
The sub-index $c$ in the upper limit of the sums indicates that these are carried out only over those $j$-values with $r_{ij}$ being in class $c$. Translating the previous equations into matrix notation, one gets the following expression,
\begin{equation}
\mathbb{F}_\alpha = \mathbb{A}_\alpha \mathbb{G} + \mathbb{D}_\alpha.
\label{matrix}
\end{equation}
The matrices $\mathbb{F}_\alpha$ and $\mathbb{D}_\alpha$ have only one column, with $\mathcal{N} N_l$ rows, where $\mathcal{N}$ is the number of simulated configurations used to gather statistics. Typical values for this product are around $10^{6} \sim 10^7$. The matrix $\mathbb{G}$ has one column with $C$ rows, typically about $10^{2}$. The matrices $\mathbb{A}_\alpha$ are arrays with the same number of rows as $\mathbb{F}_\alpha$, and $C$ columns. It should be mentioned that the minimization procedure used in this scheme allows the intermediate calculation of some matrix products, so that we do not actually need to keep in memory more than a few arrays of size $C\times C$.

Linear equations represented in Eq. (\ref{matrix}) are highly over-determined, since there are many more rows than columns. To overcome this difficulty, we use the LS method~\cite{press_1992}, asking that the values of the elements of $\mathbb{G}$' be minimized from those of $\mathbb{D}^\mathtt{T}_\alpha \mathbb{D}_\alpha$. This leads to the equation,
\begin{equation}
	\mathbb{A}^\mathtt{T}_\alpha \mathbb{A}_\alpha \mathbb{G} = \mathbb{A}^\mathtt{T}_\alpha \mathbb{F}_\alpha,
	\label{matrix3}
\end{equation}
which represents a closed and well-defined system of $C$ linear equations with $C$ variables. They are solved using Singular Value Decomposition (SVD), since the matrices $\mathbb{A}^\mathtt{T}_\alpha \mathbb{A}_\alpha$ are in general singular~\cite{press_1992}, in particular because there is no signal for $r < r_{\text{min}}$, where $r_{\text{min}}$ is the distance of the closest approach between large particles achieved in the simulation. We carry out the same procedure for all $\mathbb{F}_\alpha$, with $\alpha=x,y,z$ independently, so we get three different solutions for $\mathbb{G}$. A comparison between them provides some validation for our approach; in all cases, we found them to be equal within the statistical fluctuations. The values shown later are the averages of the three solutions.

Recently, this formalism has been explicitly used to better understand the depletion forces beyond the diluted limit of large species \cite{delossantos-lopez-PA-2024} and to discuss the role of the entropic gate in stabilizing colloidal dispersions driven by purely entropic contributions \cite{delossantos-lopez_JCP_2021}.


\subsection{Integral equations theory for effective interactions}

In the following, we refer to the integral equations theory for effective interactions in complex fluids \cite{alcaraz-klein,ramon-ro-al-2,delossantos-lopez_JCP_2021}, which takes advantage of the covariance of the Ornstein-Zernike equation under contractions of the description. For homogeneous mixtures that contain $p$ species of spherical particles, the Ornstein-Zernike (OZ) equation, written in Fourier space, is given by \citep{Hansen1986},
\begin{equation}
    \tilde{h}_{ij}(q)=\tilde{c}_{ij}(q)+\sum_{k=1}^{p}n_{k} \tilde{h}_{ik}(q)\tilde{c}_{kj}(q).
    \label{BOZ}
\end{equation}
 Here, $\tilde{h}_{ij}(q)$ is the Fourier transform of the total correlation function $h_{ij}(r)$ between particles of species $i$ and $j$, with $q$ and $r$ being the wave number and the distance between the centers of the particles, respectively. $c_{ij}(r)$ is the direct correlation function and $n_{k}$ the number density of species $k$. From the perspective of this theory, the structure of the visible species can be described with the effective potential resulting from the integration of the degrees of freedom of the remaining species, leading to the same distribution that would be obtained from a complete description of the system in terms of the bare potentials among all species \cite{alcaraz-klein,ramon-ro-al-2,delossantos-lopez_JCP_2021}.

When this theoretical approach is used in binary mixtures of large ($l$) and small ($s$) spheres, the resulting depletion potential between the former ones, $\beta u^{\mathrm{eff}}_{ll}(r)$, after the contraction of the latter ones, can be written as \cite{alcaraz-klein,ramon-ro-al-2}
\begin{eqnarray}
\beta u^{\mathrm{eff}}_{ll}(r) & = & \beta u_{ll}(r)+ \nonumber \\ & & [c_{ll}(r)-c^{\mathrm{eff}}_{ll}(r)]+
[b^{\mathrm{eff}}_{ll}(r)-b_{ll}(r)],
\label{dp1}    
\end{eqnarray}
or alternatively,
\begin{eqnarray}
\beta u^{\mathrm{eff}}_{ll}(r) & = & \beta w_{ll}(r)+ \nonumber \\ & & n_l \int_V c^{\mathrm{eff}}_{ll}(r) h_{ll}(|\bm{r}-\bm{r}'|) d\bm{r}'+ b^\mathrm{eff}_{ll}(r),
\label{dp2}       
\end{eqnarray}
with
\begin{equation}
\tilde{c}^\mathrm{eff}_{ll}(q)=\tilde{c}_{ll}(q)+\frac{\tilde{c}_{ls}(q) n_s \tilde{c}_{sl}(q)}{1-n_s \tilde{c}_{ss}(q)},
\label{ceff}       
\end{equation}
being the effective direct correlation function between large particles. Here, $\beta=1/k_B T$ is the inverse of the thermal energy, respectively, with $k_{B}$ being the Boltzmann constant and $T$ the absolute temperature. The bare interaction potential between the particles of species $i$ and $j$ in the mixture ($i,j=s,l$) is $u_{ij}(r)$, while $b_{ij}(r)$ is the bridge correlations \cite{Hansen1986}. The superscript $\mathrm{eff}$ denotes the corresponding effective functions (after the contraction of the description).  The spatial distribution of large particles must be the same at both levels of description, i.e., $h^\mathrm{eff}_{ll}(r)=h_{ll}(r)$. The potential of mean force between large particles is $\beta w_{ll}(r)=-\ln g_{ll}(r)$, with $g_{ll}(r)=h_{ll}(r)+1$ being the corresponding radial distribution function \cite{Hansen1986}.

Equation (\ref{dp1}), together with Eq. (\ref{ceff}), highlights the fact that the effective potential among large spheres emerges from the particle correlations, i.e., the depletion potential is then given by the bare one plus terms depending on the correlations between large particles mediated by the small ones. An interesting and natural result obtained from Eq. (\ref{dp2}), and confirmed by independent experiments and molecular simulations \cite{perera}, is that the effective potential becomes the potential of mean force when only a few large particles are present, that is, in the dilute limit of the large species (the leading term in $b^\mathrm{eff}_{ll}(r)$ is quadratic in $n_l$ \cite{Hansen1986}), even for high concentrations of small particles.

The correlation terms in Eq.~(\ref{dp1}) have been evaluated under many approximations, and the predictions of this theoretical framework have been systematically tested for self-consistency by ensuring the equality of the radial distribution functions obtained by both routes, simulating the complete mixture with the bare potentials and the effective monodisperse system of large particles with the depletion interaction~\cite{erik,perera}.

To numerically evaluate Eq. (\ref{BOZ}), a five-parameter version of the Ng method \citep{NgMethod} is used to solve the coupled OZ equations closed with the modified-Verlet closure relation, which is given by \citep{erik},
\begin{equation}
b_{ij}(r)=-0.5\frac{\gamma_{ij}^{2}(r)}{1+0.8{\gamma_{ij}(r)}},
\label{BmV}
\end{equation}
where $\gamma_{ij}(r)=h_{ij}(r)-c_{ij}(r)$. It has been shown that Eq. (\ref{BmV}) is an accurate closure relation for colloidal dispersions interacting with short-range repulsive potentials \cite{erik,perera,delossantos-lopez_JCP_2021,delossantos-lopez_JCP_2022,delossantos-lopez-PA-2024,Bedolla25}.

\section{Polynomial fits for depletion force}
\label{polynomial_fit}

This appendix provides the coefficients obtained from polynomial fits of the depletion force at different packing fractions, $\phi_l$ and $\phi_s^{\text{res}}$. These fits can be employed to interpolate or extrapolate the depletion potential at sets of packing fractions not explicitly reported in the main text.

\begin{table*}[b]
\caption{\label{tableofcoefficients}
Coefficients for the polynomial fit $f_d = A + By + Cy^2 + Dy^3$, where $y = r/\sigma_{l} - 1$. The shift in $r/\sigma_{l}$ is done only for convenience. The error bars (not shown) are minimal for $A$ and $B$, but can be large for the other two coefficients.}
\begin{ruledtabular}
\begin{tabular}{c|ccccc}
$\phi_s^{\text{res}}$ & $\phi_l$ & $A$ & $B$ & $C$ & $D$ \\
\hline
0.025 & 0.1 & -0.1884 & 0.3247 & 0.3227 & -0.1958 \\
      & 0.2 & -0.1773 & 0.3195 & 0.1597 &  0.0775 \\
      & 0.3 & -0.1588 & 0.2851 & 0.1693 &  0.0127 \\
      & 0.4 & -0.1322 & 0.2347 & 0.1605 & -0.0090 \\
\hline
0.050 & 0.1 & -0.3974 & 0.7834 & 0.2181 &  0.2801 \\
      & 0.2 & -0.3746 & 0.7264 & 0.2627 &  0.1879 \\
      & 0.3 & -0.3409 & 0.6098 & 0.4768 & -0.1348 \\
      & 0.4 & -0.2897 & 0.5108 & 0.4619 & -0.2231 \\
\hline
0.075 & 0.1 & -0.6156 & 1.1656 & 0.7868 & -0.0433 \\
      & 0.2 & -0.5909 & 1.1310 & 0.7861 & -0.3504 \\
      & 0.3 & -0.5405 & 0.9752 & 0.9155 & -0.5022 \\
      & 0.4 & -0.4731 & 0.8728 & 0.6214 & -0.1289 \\
\hline
0.100 & 0.1 & -0.8642 & 1.7062 & 1.4564 & -1.1130 \\
      & 0.2 & -0.8247 & 1.6265 & 1.2391 & -0.7269 \\
      & 0.3 & -0.7679 & 1.4977 & 0.9612 & -0.2298 \\
      & 0.4 & -0.6810 & 1.2637 & 1.0744 & -0.4677
\end{tabular}
\end{ruledtabular}
\end{table*}

\bibliography{depletion-high-order}

\end{document}